\documentclass[twocolumn,showpacs]{revtex4-1}

\usepackage{color}
\usepackage{graphicx}

\begin{document}  

\title{Many-body effects on Landau-level spectra and cyclotron resonance in graphene }
\author{K. Shizuya}
\affiliation{Yukawa Institute for Theoretical Physics\\
Kyoto University,~Kyoto 606-8502,~Japan}

\begin{abstract} 
Recently Russell {\it et al.} [Phys.~Rev.~Lett.~{\bf 120}, 047401 (2018)] 
have reported a clear signal of many-particle contributions 
to cyclotron resonance in high-mobility hBN-encapsulated graphene, 
observing significant variations of resonance energies as a function of the filling factor $\nu$ 
for a series of interband channels.
To elucidate their results, 
Coulombic contributions to the Landau-level spectra and cyclotron resonance in graphene are examined 
with a possible band gap taken into account
and with emphasis on revealing electron-hole ($eh$) conjugation symmetry 
underlying such level and resonance spectra.  
Theory, based on the single-mode approximation, 
gives a practically good account of the experimental data;
the data suggest a band gap of $\sim$ 10 meV and 
show a profile that apparently reflects $eh$ conjugation symmetry.

\end{abstract}


\maketitle

\section{Introduction}

 Graphene supports as charge carriers  massless Dirac fermions~\cite{NG,ZTSK,GN_rev}, 
whose spinor nature derives from the underlying honey-comb lattice structure. 
In a magnetic field, graphene reveals its $\lq\lq$relativistic" character,
leading to a particle-hole symmetric and unequally-spaced
tower of Landau levels, along with some characteristic zero-energy levels.
It gives rise to a variety of cyclotron resonance (CR) channels~\cite{AbF}, 
both intraband and interband.
This is in sharp contrast to standard quantum Hall systems 
(with quadratic energy dispersion),
in which CR takes place only 
between each adjacent pair of evenly-spaced Landau levels,
hence at a single frequency $\omega_{c}= eB/m^{*}$, 
which, according to Kohn's theorem~\cite{Kohn,KH}, 
is unaffected by electron-electron interactions. 
CR in graphene and related Dirac-electron systems thus provides 
an ideal ground for exploring many-body effects.

Actually, graphene is an intrinsically many-body system of electrons equipped 
with the valence band acting as the Dirac sea.
Quantum fluctuations of the Dirac sea are generally sizable, leading to ultraviolet divergences,
and one has to go through renormalization properly to extract observable many-body effects, 
such as velocity renormalization~\cite{GGV}, 
Coulombic corrections to CR~\cite{IWF,BM,KS_CR},
and collective excitations~\cite{RFG}.

Experiments have already explored, via infrared spectroscopy, 
some basic features of the Landau-level spectra and associated CR
in monolayer~\cite{JHTWS,DCNN,HCJL}
and bilayer~\cite{MNEB,HJTS,OFBB} graphene.
Coulombic corrections to CR escaped detection in an early experiment~\cite{SMPB},
and were first observed (in a sample of graphene on Si/SiO$_{2}$)
via the comparison of a certain set of  intra- and interband transitions~\cite{JHTWS}.
The running of the Fermi velocity $v$ 
under a change in scale
was also observed~\cite{EGMM,FBNH}.

Meanwhile high-mobility samples became available such as suspended graphene and 
graphene on hexagonal boron nitride (hBN).
In particular, the graphene/hBN device attracts attention for its flatness and 
a possible opening of a band gap~\cite{HuntYY,CSYL,WoodsBE}
due to a small lattice mismatch and weak interlayer interaction.

Recently Russell {\it et al}.~\cite{RZTW} have reported 
a direct signal of many-particle contributions to CR 
in high-mobility hBN-encapsulated monolayer graphene. 
They observed significant variations of resonance energies 
over a certain range of filling factor $\nu$ under fixed magnetic field $B$.

The purpose of this paper is to examine the Coulombic contributions 
to Landau-level and CR spectra in graphene, 
with a possible band gap taken into account, 
and to interpret the observed data of Ref.~\cite{RZTW}.
In a magnetic field a nonzero band gap requires careful handling 
by renormalization with counterterms nonlinear in the band gap, 
which fortunately are determined by referring to the theory in free space.  
In our analysis, particular attention is paid to the
electron-hole ($eh$) conjugation symmetry 
intrinsic to the basic effective Hamiltonian for graphene.
We clarify how it governs the Landau-level and CR spectra and note 
that it is indeed well reflected in the observed data. 
The theory, based on the single-mode approximation (SMA)~\cite{MOG,GMP,MZ,KS_sma}, 
gives a practically good account of the experimental data: 
The zero-mode Landau levels ($n=0_{\pm}$) are particularly sensitive to the band gap 
and a close look into the related data suggests a band gap of 
$\sim$ 10 meV, while the strength of the Coulomb potential is estimated 
from relative variations in resonance energy for a series of interband channels.

    In Secs.~II and III we review the effective theory of graphene in a magnetic field 
and some basic features of Coulombic corrections.
In Sec.~IV we examine the detailed structure of the level and CR spectra 
and reveal the underlying $eh$-conjugation symmetry.
In Sec.~V we carry out renormalization and examine
how the Coulomb-corrected level and CR spectra change with the filling of levels. 
In Secs.~VI and VII we take a close look into each specific behavior 
of leading sets of interband CR channels and compare them with the observed data. 
Section~VIII is devoted to a summary and discussion.

\section{graphene}

The electrons in graphene are described by two-component spinors 
on two inequivalent lattice sites. 
They acquire a linear spectrum (with velocity $ v \sim 10^{6}\, $m/s) 
near the two inequivalent Fermi points $(K,K')$ in momentum space, 
and are described by an effective Hamiltonian of the form~\cite{Semenoff}
\begin{eqnarray} 
H &=&\int dx dy \, \{ \Psi^{\dag}_{+} {\cal H}_{+}\Psi_{+} + \Psi_{-}^{\dag} {\cal H}_{-}\Psi_{-} \},  
\nonumber\\
{\cal H}_{\pm} &=& 
v\, (\Pi_{1}\sigma^{1}+ \Pi_{2}\sigma^{2}) \pm M\, \sigma^{3},
\label{H_GR}
\end{eqnarray}
where $\Pi_{i}= p_{i}+eA_{i}$  [with $i=(1,2)$ or $(x,y)$] 
involve coupling to potentials $A_{i}$ 
and $\sigma^{i}$ denote Pauli matrices.
The Hamiltonians ${\cal H}_{\pm}$ describe electrons 
at two different valleys $a \in (K,K')$ per spin, 
and $M$ stands for a possible valley gap; 
we take $M > 0$, without loss of generality.

Let us place graphene in a uniform magnetic field $B_{z}=B>0$ 
by setting $A_{i}=(-By,0)$.
The electron spectrum then forms an infinite tower of 
Landau levels of energy 
\begin{equation} 
\epsilon_{n} =s_{n}\,  \omega_{c} \sqrt{|n|+\mu^{2}}
\end{equation}
at each valley (with $s_{n}\equiv  {\rm sgn}[n] = \pm1$), 
labeled by integers $n \in (0,\pm 1, \pm2, \dots)$ and
$p_{x}$, of which only the $n=0$ levels split in valley
(hence to be denoted as $n=0_{\pm}$), 
\begin{equation}
\epsilon_{0_{\mp}}= \mp  M = \mp \omega_{c}\,  \mu   \ \ {\rm for}\ K/K'.
\end{equation}
Here we have set, along with magnetic length $\ell \equiv 1/\sqrt{eB}$, 
\begin{equation}
\omega_{c}\equiv \sqrt{2}\, v/\ell 
\approx 36.3 \times v(10^{6}{\rm m/s})\, \sqrt{B({\rm T})}\ {\rm meV},
\end{equation}
and $\mu\equiv M/\omega_{c}$.
Thus, for each integer $|n| \equiv N$,
there are in general two modes with $n=\pm N$ 
(of positive/negative energy) at each valley per spin,
apart from the $n=0_{\pm}$ modes.

The eigenmodes of ${\cal H}_{+}$ at valley $K$ are written as 
\begin{equation}
\phi_{n} = \big( |N-1 \rangle\, b_{n}, |N \rangle\, c_{n} \big)^{\rm t}\ \ 
{\rm and}\ N \equiv |n|
\label{psi_n}
\end{equation}
[here only the orbital eigenmodes are shown using the harmonic-oscillator basis 
$\{ |N\rangle\}$],
with $(b_{n}, c_{n})$ given by
\begin{eqnarray} 
&&(b_{n}, c_{n})= \textstyle{1\over{\sqrt{2}}}\, \left( \sqrt{1+ \mu/e_{n} }, -s_{n} \sqrt{1- \mu/e_{n} }\right),
\nonumber\\
&&(b_{0_{-}}, c_{0_{-}}) = (0, 1),    
\label{bc_muzero}
\end{eqnarray}
where 
$e_{n} = e_{n}(\mu) \equiv \epsilon_{n}/\omega_{c}= s_{n}\sqrt{N + \mu^2}$.

One can pass to another valley $K'$ by simply setting $\mu\rightarrow - \mu$
in the $K$-valley expressions. 
Alternatively, note the relation $\sigma^{3}\, {\cal H}_{-}\sigma^{3} = -{\cal H}_{+}$,
which implies that 
the spectra and eigenmodes of valley $K'$ are determined by those of valley $K$,
\begin{eqnarray}
\phi_{n}|^{K'} &=& \sigma^{3}\, \phi_{-n}|^{K},  \nonumber\\
\epsilon_{n}|^{K'} &=& - \epsilon_{-n}|^{K},\  
(b_{n},  c_{n})|^{K'}=(b_{-n}, -c_{-n})|^{K}.
\label{KandKprime}
\end{eqnarray}
This represents the basic invariance of $H$ under electron-hole ($eh$) conjugation,
 i.e., forming another valley by interchanging the electron and hole bands in a valley.
One can also  define $eh$ conjugation within each valley by replacing 
$\mu\rightarrow -\mu$,
\begin{equation}
(\epsilon_{n}, b_{n},  c_{n})=(- \epsilon_{-n}, b_{-n}, -c_{-n})|^{\mu\rightarrow-\mu},
\label{eh_KandK}
\end{equation}
in obvious notation,  with $n=0 \rightarrow 0_{\mp}$ in each valley.

The Landau-level structure is made explicit by passing to
the $|n,y_{0}\rangle$ basis  (with $y_{0}\equiv \ell^{2}p_{x}$) 
and the field $\psi^{n;a}_{\alpha}(y_{0})$, 
where $n$ refers to the Landau level, 
$a \in (K,K')$ to the valley and
$\alpha \in (\downarrow, \uparrow)$ 
to the spin.
The Lagrangian thereby reads 
\begin{equation}
L= \int dy_{0} \sum_{n}\sum_{a,\alpha}
(\psi^{n;a}_{\alpha})^{\dag}(i\partial_{t} -\epsilon_{n}^{a}) \psi^{n;a}_{\alpha}
\label{Lag_onebody}
\end{equation}
and the charge density 
$\rho_{-{\bf p}} =\int d^{2}{\bf x}\,  e^{i {\bf p\cdot x}}\,\rho$ 
with $\rho = \Psi_{+}^{\dag} \Psi_{+} +   \Psi_{-}^{\dag} \Psi_{-}$ 
is written as~\cite{KS_screening}
\begin{eqnarray}
\rho_{-{\bf p}} &=& \gamma_{\bf p}\sum_{m, n =-\infty}^{\infty}
\sum_{a,\alpha} g^{m n;a}_{\bf p}\, 
R^{m n;aa}_{\alpha\alpha; -{\bf p}}, \nonumber\\
R^{mn;ab}_{\alpha\beta; -{\bf p}}&\equiv& \int dy_{0}\,
{\psi^{m;a}_{\alpha}}^{\dag}(y_{0})\, e^{i{\bf p\cdot r}}\,
\psi^{n;b}_{\beta} (y_{0}),
\label{chargeoperator}
\end{eqnarray}
with $\gamma_{\bf p} =  e^{- \ell^{2} {\bf p}^{2}/4}$. 
Here, ${\bf r} = (i\ell^{2}\partial/\partial y_{0}, y_{0})$
stands for the center coordinate with uncertainty 
$[r_{x}, r_{y}] =i\ell^{2}$. 
The charge operators $R^{mn;ab}_{\alpha\beta; -{\bf p}}$ obey 
the $W_{\infty}$ algebra~\cite{GMP} that reflects this uncertainty.

The coefficient matrix $ g^{m n;a}_{\bf p}= g^{m n}_{\bf p}|^{a}$ 
at valley $a$ is given by
\begin{equation}
g^{m n;a}_{\bf p} = b_{m}^{a}\, b_{n}^{a}\, f_{\bf p}^{|m|-1,|n|-1}
+ c_{m}^{a}\, c_{n}^{a}\, f_{\bf p}^{|m|,|n|}, 
\label{gkn}
\end{equation}
where $b^{a}_{n}=b_{n}|^{a}$, etc., and 
\begin{equation}
f^{m n}_{\bf p} 
= \sqrt{n!/m!}\,
({ i \ell p/\sqrt{2}} )^{m-n}\, L^{(m-n)}_{n}
(\textstyle{1\over{2}}\ell^{2}{\bf p}^{2})
\end{equation}
for $m \ge n\ge0$, and $f^{n m}_{\bf p} = (f^{m n}_{\bf -p})^{\dag}$;
$p=p_{y}\! +i\, p_{x}$; it is understood that 
$f^{mn}_{\bf p}=0$ for $m<0$ or $n<0$.
In view of Eqs.~(\ref{KandKprime}) and (\ref{eh_KandK}),  
$g^{-m,-n;a}_{\bf p}$ are related to $g^{m,n;a}_{\bf p}$ with the sign of $\mu$ reversed 
and hence to those of the other valley,
\begin{equation}
g^{-m,-n;a}_{\bf p}= g^{m,n;a}_{\bf p}|_{\mu\rightarrow -\mu}, \ g^{mn;K'}_{\bf p}= g^{-m,-n;K}_{\bf p}.
\label{gmnKvsKp}
\end{equation}
Some explicit forms of $g^{m n;a}_{\bf p}$ are 
\begin{eqnarray}
g^{00}_{\bf p} &=& 1, \ \
g^{11}_{\bf p} =1-(c_{1})^{2}\, \textstyle{1\over{2}}\ell^{2}{\bf p}^{2}, 
\nonumber\\
g^{10}_{\bf p} &=& ic_{1} \ell\, p/\sqrt{2}, \ \
g^{01}_{\bf p} = i c_{1} \ell\, p^{\dag}/\sqrt{2},
\label{example_g}
\end{eqnarray}
with 
$c_{1}|^{K}= -\sqrt{1-\mu/e_{1}}$.

From now on we frequently suppress
summations over levels $n$, valleys $a$ and spins $\alpha$, 
with the convention that the sum is taken over repeated indices.
The one-body Hamiltonian $H$ is thereby written as
\begin{equation}
H = \epsilon^{a}_{n}\, R^{nn;aa}_{\beta\beta;{\bf 0}} 
- \mu_{\rm Z}\, (\sigma^{3}/2)_{\alpha \beta} R^{nn;aa}_{\alpha\beta;{\bf 0}}. 
\label{Hzero}
\end{equation}
Here, for generality, the Zeeman term 
$\mu_{\rm Z} \equiv g^{*}\mu_{\rm B}B$
is introduced.
Actually, spin splitting is relatively weak, $\mu_{\rm Z} \approx 0.12\,  B({\rm T})$ meV.
We therefore note its presence but take no explicit account of it numerically.

The Coulomb interaction 
$V= {1\over{2}} \sum_{\bf p} v_{\bf p}\,  {:\! \rho_{\bf -p}\, \rho_{\bf p}\!:}$ 
is written as
\begin{equation}
V = {1\over{2}} \sum_{\bf p}
v_{\bf p}\,\gamma_{\bf p}^{2}\,  
g^{jk; a}_{\bf p}\, g^{m n;b}_{\bf -p}
:\! R^{j k;aa}_{\alpha\alpha;{\bf -p}}\, 
R^{m n;bb}_{\beta\beta;{\bf p}}\! : , 
\label{VCoul}
\end{equation}
with the potential $v_{\bf p}= 2\pi \alpha/(\epsilon_{\rm b} |{\bf p}|)$,
$\alpha \equiv e^{2}/(4 \pi \epsilon_{0})$ and 
the substrate dielectric constant $\epsilon_{\rm b}$;
$\sum_{\bf p} \equiv \int d^{2}{\bf p}/(2\pi)^{2}$ and we set 
$\delta_{\bf p,0}\! \equiv (2\pi)^2 \delta^{2}({\bf p})$.   
As usual, normal ordering is defined as  
$:\! R^{j k} R^{m n}\! : \ \propto (\psi^{m})^{\dag} (\psi^{j})^{\dag}  \psi^{k}\psi^{n}$.

\section{Coulombic corrections}

In this section we study the Coulombic contributions to Landau-level spectra 
and associated CR in graphene.
 Let us suppose that a uniform ground state $|{\rm Gr}\rangle$ is realized at some filling factor $\nu$ 
 in a magnetic field, with the charge expectation values 
\begin{equation} 
\langle {\rm Gr}|R^{mn;ab}_{\alpha\beta; {\bf p}}|{\rm Gr}\rangle
= \bar{\rho}\, \nu_{n}^{a; \alpha}\delta^{mn}\delta^{ab}\delta^{\alpha\beta}\, 
\delta_{\bf p,0}
\end{equation}
for good (i.e., diagonal) quantum numbers $(n, a, \alpha)$, where 
$0\le \nu_{n}^{a; \alpha} \le 1$ stands for the filling fraction 
of the $(n, a, \alpha)$ level and  $\bar{\rho}=1/(2\pi \ell^2)$.

The Coulomb direct interaction leads to a divergent self-energy 
$\propto v_{\bf p \rightarrow 0}$, which, as usual, is removed 
if one takes into account a neutralizing positive background.
The exchange interaction gives rise
to corrections to level spectra $\epsilon_{n}^{a; \alpha}$ of the form 
\begin{equation}
\Delta \epsilon_{n}^{a; \alpha}
= -\sum_{\bf p}v_{\bf p}\gamma_{\bf p}^{2}\, \sum_{m} \nu_{m}^{a; \alpha}\, |g^{n m;a}_{\bf p}|^2 ,
\label{self_En}
\end{equation}
where the sum is taken over filled levels $m$.
An exchange interaction, in calculating Coulombic corrections, preserves the spin and valley 
$(\alpha,a)$. Accordingly, from now on we suppress them
 and mainly display the $K$-valley expressions.

Let us next study CR, namely, optical interlevel transitions at zero momentum transfer,
with the selection rule~\cite{AbF} $\Delta |n|=\pm 1$ for graphene, 
 i.e., 
(i) intraband channels $n\leftarrow n-1$ and  $-(n-1) \leftarrow -n$
and 
(ii) interband channels $n\leftarrow -(n-1)$ and  $(n-1) \leftarrow -n$ for $n=1,2,3, \cdots$.
Interband CR is specific to Dirac electrons and takes place over a certain range of filling factor $\nu$.
Consider now CR from level $j$ to level $n$ for each (valley, spin)=$(a,\alpha)$ channel
and denote the associated excitation energy as
\begin{eqnarray}
\epsilon_{\rm exc}^{n\leftarrow j} &=& \epsilon_{n}- \epsilon_{j} +\Delta \epsilon^{n,j}.
\end{eqnarray}
The mean-field treatment, such as the SMA,
leads to Coulombic corrections of the form~\cite{KS_CR,MOG,GMP,MZ,KS_sma} 
\begin{equation}
\Delta \epsilon^{n,j}= \Delta \epsilon_{n} -\Delta \epsilon_{j} 
- (\nu_{j}-  \nu_{n} )\, \sum_{\bf p} v_{\bf p}\gamma_{\bf p}^{2}\, g^{nn}_{\bf -p}\, g^{jj}_{\bf p} 
\label{Eexc_SMA} 
\end{equation}
for each (valley, spin) channel; 
see Ref.~[28] for a refined formulation of SMA calculations and a derivation of Eq.~(\ref{Eexc_SMA}).
The corrections $\Delta \epsilon^{n,j}$ thus consist of self-energies 
$(\Delta \epsilon_{n},\Delta \epsilon_{j})$ [in Eq.~(\ref{self_En})] 
of the excited electron and created hole and the Coulomb attraction 
$\propto \sum_{\bf p} v_{\bf p}\gamma_{\bf p}^{2}\, g^{nn}_{\bf -p}\, g^{jj}_{\bf p}$ between them.

Actually Eq.~(\ref{Eexc_SMA}) is an expression adequate for integer filling of the ground state. 
When the initial or final level is only partially filled, $\Delta \epsilon^{n,j}$ acquires
an extra contribution from nontrivial correlations within such a level, 
as characterized, in the SMA~\cite{MOG,GMP},    
by the static structure factor $\hat{s}_{n}({\bf p})$ 
in the projected structure function
$\langle {\rm Gr}| R^{nn}_{\bf -p} R^{nn}_{\bf p}|{\rm Gr} \rangle
= |\langle  {\rm Gr}| R^{nn}_{\bf p}| {\rm Gr}\rangle|^2
+ \delta_{\bf 0,0} \bar{\rho}\, \nu_{n}\, \hat{s}_{n}({\bf p})$ (for fixed $n$);
$\delta_{\bf 0,0}= \int d^2{\bf x}$.
In general, $\hat{s}_{n}({\bf p}) \rightarrow 0$ for a filled level. 
The Hartree-Fock approximation~\cite{MOG} yields $\hat{s}_{n}({\bf p})=1-\nu_{n}$, 
and Eq.~(\ref{Eexc_SMA}) is actually a Hartree-Fock expression
with this choice of  $\hat{s}_{n}({\bf p})$.
In what follows we focus on the ground states of integer filling.

A remark is in order on a special feature of the $1\leftarrow 0$  resonance.
The SMA leads to a correction of the form
\begin{eqnarray}
\Delta \epsilon^{1,0}&=& -\sum_{\bf p} v_{\bf p}\gamma_{\bf p}^{2}
\Big[  \sum_{r\le -1} \{ |g^{1 r}_{\bf p}|^{2} -|g^{0r}_{\bf p}|^{2} \}
\nonumber\\
&&+\{1 - \hat{s}_{0}({\bf p})\}\, 
(|g^{1 0}_{\bf p}|^{2} -|g^{00}_{\bf p}|^{2} + g^{00}_{\bf p}\, g^{11}_{\bf -p}) \Big], 
\ \ \ \ \ 
\label{DEone-zero}
\end{eqnarray}
when the structure factor $\hat{s}_{0}({\bf p})$ of the $0_{-}$ level is retained. 
For conventional electrons with quadratic dispersion
one only has the last term $\propto \{1 - \hat{s}_{0}({\bf p})\}$, 
though it actually vanishes in accordance with Kohn's theorem~\cite{Kohn}. 
It happens to vanish also for this $\Delta \epsilon^{1,0}$ of graphene, 
since 
$|g^{1 0}_{\bf p}|^{2} -|g^{00}_{\bf p}|^{2} + g^{00}_{\bf p}\, g^{11}_{\bf -p}=0$ holds,
as one can verify using Eq.~(\ref{example_g}).
Thus, in the SMA, $\Delta \epsilon^{1,0}$ consists solely of self-energy corrections 
due to the filled valence band
and is actually logarithmically divergent.

\section{electron-hole conjugation}

The self-energies $\Delta \epsilon_{n}$ 
involve a sum over infinitely many filled levels in the valence band.  
Their structure is better clarified if one notes  
the completeness relation~\cite{fn_comp}
\begin{equation}
\sum_{k=-\infty}^{\infty}|g^{n k}_{\bf p}|^2 
= e^{{1\over{2}}\ell^2 {\bf p}^2}
=1/ \gamma_{\bf p}^{2}.
\label{completeness-rel}
\end{equation}
The half-infinite sum in $\Delta \epsilon_{n}$ is thereby rewritten as
\begin{eqnarray}
\gamma_{\bf p}^{2}\sum_{k\le -1}|g^{n k}_{\bf p}|^2
&=&\textstyle {1\over{2}} -   {1\over{2}}\,  {\cal F}_{n}(z;\mu) 
- {1\over{2}}\, \gamma_{\bf p}^{2}|g^{n 0_{-}}_{\bf p}|^2, 
\nonumber\\
{\cal F}_{n}(z;\mu) &\equiv&
\gamma_{\bf p}^{2}\sum_{k=1}^{\infty}\{ |g^{nk}_{\bf p}|^2 -  |g^{n,-k}_{\bf p}|^2 \},
\end{eqnarray}
where 
$z= {1\over{2}}\ell^2 {\bf p}^{2}$.
In particular, 
\begin{eqnarray}
{\cal F}_{n}(z; 0) &\stackrel{n\ge 1}{=}&
e^{-z}\sum_{k=1}^{\infty} \sqrt{{k\over{n}}}\, {n!\over{k!}}\, 
z^{k-n} L^{k-n}_{n-1}(z)\,  L^{k-n}_{n}(z),
\nonumber\\
{\cal F}_{0_{-}}(z;\mu) 
&=&  -\mu \sum_{k=1}^{\infty}{1\over{e_{k}}}\, {z^{k}\over{k!}}\, e^{-z},
\label{Fnz}
\end{eqnarray}
with $e_{k}= \sqrt{k + \mu^2}$ for $k>0$.

The self-energies $\Delta \epsilon_{n}$ are now rewritten as 
\begin{equation}
 \Delta \epsilon_{n}  = \sum_{\bf p}v_{\bf p}\,\Big[
- {\textstyle{1\over{2}}} +  {\textstyle{1\over{2}}}{\cal F}_{n}(z;\mu)
- \sum_{k} \nu [k]\,  \gamma_{\bf p}^{2}  |g^{nk}_{\bf p}|^2\Big].
\label{DEn_selfenergy}
\end{equation}
Here the last term with the $\lq\lq$electron-hole" filling factor,
\begin{equation}
\nu[k] = \nu_{k}\, \theta_{(k \ge1)} -  (1-\nu_{k})\theta_{(k\le-1)}
 + (\nu_{0}- {\textstyle{1\over{2}} })\, \delta_{k,0},
 \label{nu_k_ehsym}
 \end{equation}
where 
$\theta_{(k\ge1)} =1$ for $k\ge1$ and $\theta_{(k\ge1)} =0$ otherwise, etc.,
stands for contributions from a finite number of 
filled electron or hole levels around the $n=0$ level.
The filled valence band has led to 
corrections $\propto - {\textstyle{1\over{2}}} +  {\textstyle{1\over{2}}}{\cal F}_{n}(z;\mu)$,
of which the $-{1\over{2}}$ term, common to all levels $n$, is safely eliminated 
by adjusting zero of energy.
${\cal F}_{n}(z;\mu)$ thus represent genuine many-body corrections.
In view of Eq.~(\ref{gmnKvsKp}), ${\cal F}_{\pm n}$ are related 
in each valley or between the valleys,
\begin{eqnarray}
{\cal F}_{-n}(z;\mu) &=& - {\cal F}_{n}(z; -\mu),
\nonumber\\
{\cal F}_{n}(z;\mu)|^{K'} &=& - {\cal F}_{-n}(z;\mu)|^{K} =   {\cal F}_{n}(z;-\mu)|^{K}.
\label{eh_sym_Fn}
\end{eqnarray}

Let us now disclose a key property of $\nu[k]$ defined in Eq.~(\ref{nu_k_ehsym}):
$\nu[-k]$ equals $-\nu[k]$ with each $\nu_{k}$    
replaced by $(1- \nu_{-k})$ [and $\nu_{0_{-}} \rightarrow 1- \nu_{0_{-}}$] in the latter. 
This means that $\nu[k]$ changes sign upon interchanging  
the electron and hole bands according to $\nu_{k} \rightarrow 1- \nu_{-k}$, i.e., 
via $eh$ conjugation. 
Noting Eqs.~(\ref{gmnKvsKp}) and~(\ref{eh_sym_Fn}) then allows one to relate
 $\Delta \epsilon_{n} =  \sum_{\bf p}v_{\bf p}\,[ {1\over{2}}{\cal F}_{n}(z;\mu)+ \cdots]$ 
to $-\Delta \epsilon_{-n}$
in the same valley or in another valley. The result is
\begin{eqnarray}
\Delta \epsilon_{n} &=& 
-\Delta \epsilon_{-n}|_{\nu_{k}\rightarrow1-\nu_{-k}}^{\mu\rightarrow -\mu}
= -\Delta \epsilon_{-n}|_{\nu_{k}\rightarrow1-\nu_{-k}}^{K'},
\label{EnToEmn}
\end{eqnarray}
in obvious notation.
To make the situation clearer, 
let us imagine valley $K$ filled up to  level $n_{\rm f} =m$,
i.e., $\nu_{k}=1$ for $k\le m$;  $n_{\rm f}$ specifies the uppermost filled level.
Interchanging electrons and holes yields a configuration with levels filled up to $n_{\rm f} =-m-1$.
Thus, via $eh$ conjugation, valley $K$ with $n_{\rm f} =m$ turns into valley $K'$ with  
$n_{\rm f} =-m-1$, and vice versa. 
One can now rewrite Eq.~(\ref{EnToEmn}) for the full spectra 
$\hat{\epsilon}_{n} = \epsilon_{n} + \Delta  \epsilon_{n}$ 
and denote 
\begin{equation}
\hat{\epsilon}_{n}^{K}|_{n_{\rm f}=m} =
-\hat{\epsilon}_{-n}^{K}|_{n_{\rm f}=-m-1}^{\mu\rightarrow -\mu}
= -\hat{\epsilon}_{-n}^{K'}|_{n_{\rm f}=-m-1}.
\label{En-ehconj}
\end{equation}

An analogous operation applied to Coulombic corrections 
$\Delta \epsilon^{n,j}$ in Eq.~(\ref{Eexc_SMA}) reveals that, via $eh$ conjugation,
$\Delta \epsilon^{n,j}$ turns into $\Delta \epsilon^{-j,-n}$ in another valley. 
Accordingly, the full CR spectra 
$\epsilon_{\rm exc}^{n\leftarrow -j}$ enjoy the property
 \begin{equation}
\epsilon_{\rm exc}^{n \leftarrow -j}|_{n_{\rm f}=m}^{K} \!
=\epsilon_{\rm exc}^{j \leftarrow -n}|_{n_{\rm f}=-m-1}^{K;\mu\rightarrow -\mu}
= \epsilon_{\rm exc}^{j \leftarrow -n}|^{K'}_{n_{\rm f}=-m-1}.
\label{Eexc-ehconj}
\end{equation}
Note also that, under the same $n_{\rm f}$, one can pass to another valley 
by simply reversing $\mu \rightarrow-\mu$,
\begin{equation}
\hat{\epsilon}^{K}_{n}|_{n_{\rm f}} = \hat{\epsilon}^{K'}_{n}|_{n_{\rm f}}^{\mu\rightarrow-\mu},\ \ 
\epsilon_{\rm exc}^{n \leftarrow -j}|^{K}_{n_{\rm f}} 
=\epsilon_{\rm exc}^{n \leftarrow -j}|_{n_{\rm f}}^{K'; \mu\rightarrow -\mu}.
\label{KKp_viamu}
\end{equation}

From Eq.~(\ref{Eexc-ehconj}) we learn that CR channels 
$(n\leftarrow -j)$ and $(j \leftarrow -n)$ 
form $\lq\lq eh$-conjugate" channels, interchangeable via $eh$ conjugation.
In view of the selection rule, conjugate channels of interest are $n \leftarrow-(n-1)$ 
and $n-1 \leftarrow - n$ for each $n=1, 2,\dots$, which we denote as $T_{n}$;
$T_{1} = \{ 1\! \leftarrow\! 0,\  0\! \leftarrow\! -1\}$ (with $0 \rightarrow 0_{\pm}$ at each valley),
$T_{2} = \{ 2 \leftarrow -1,\  1 \leftarrow -2\}$, $T_{3} = \{ 3 \leftarrow -2,\  2 \leftarrow -3\}$, etc.
Experimentally  signals from the conjugate channels of a given set $T_{n}$ 
are observable over a certain range of filling factor $\nu$ and are indistinguishable
unless polarized light is used.
[In contrast, $eh$-conjugate intraband channels 
$n \leftarrow (n-1)$ and $-(n-1) \leftarrow -n$ are not simultaneously observable.]

In equilibrium, the neutral ground state with total filling $\nu=0$
has the valley content  $(n_{\rm f}^{K}, n_{\rm f}^{K'})   = (0,-1)$,
which, via $eh$ conjugation, turns into $(n_{\rm f}^{K'},n_{\rm f}^{K})   = (-1,0)$.
Thus the $\nu=0$ state is $eh$-selfconjugate.
[See, in this connection, Fig.~4(a) shown later.]
In general, via $eh$ conjugation, 
a state with total filling $\nu$ turns into the state with filling $-\nu$~\cite{fn_ehconj}.
The $\nu = \pm \nu_{\rm t}$ conjugate states have essentially the same level spectra 
(up to sign and $K\leftrightarrow K'$)
\begin{equation}
(\hat{\epsilon}_{n}^{K}, \hat{\epsilon}_{n}^{K'})|_{\nu=\nu_{\rm t}} 
= (-\hat{\epsilon}_{-n}^{K'}, -\hat{\epsilon}_{-n}^{K})|_{\nu=- \nu_{\rm t}}.
\end{equation}
They also share the same excitation spectra. 
For $T_{2}$, e.g., one can write
\begin{equation}
(\epsilon_{\rm exc}^{2\leftarrow -1}, \epsilon_{\rm exc}^{1 \leftarrow -2})|^{K}_{\nu=\nu_{\rm t}} 
=(\epsilon_{\rm exc}^{1 \leftarrow -2}, \epsilon_{\rm exc}^{2\leftarrow -1})|^{K'}_{\nu= - \nu_{\rm t}},
\label{Eex-conj}
\end{equation}
i.e., the CR spectra of $T_{2}$ at one valley and total filling $\nu_{\rm t}$ are the same as 
the (conjugated) spectra at another valley and filling $-\nu_{\rm t}$.
As a result, the full ($K+K'$) spectra of $T_{2}$, now consisting of  
$(\epsilon_{\rm exc}^{2\leftarrow -1}, \epsilon_{\rm exc}^{1 \leftarrow -2})|^{K}_{\nu=\nu_{\rm t}}$
and $(\epsilon_{\rm exc}^{1 \leftarrow -2},\epsilon_{\rm exc}^{2\leftarrow -1})|^{K}_{\nu=-\nu_{\rm t}} $,
are identical at filling $\pm \nu_{\rm t}$.
This leads to a somewhat nontrivial consequence: 
The full excitation spectra of each $T_{n}$, when observed under fixed field $B$ 
over a certain range of $\nu$,
take a profile symmetric in $\nu$.
Actually, $\nu=0$ turns out to be a maximum,
as we will see later (in Fig.~2).

\section{Renormalization}

For graphene, self-energies $\Delta \epsilon_{n}$ are afflicted with ultraviolet divergences. 
In this section we study how to extract physically observable information 
out of them via renormalization.
Actually, for $n\ge 1$,
${\cal F}_{n}(z; 0)  \approx \sqrt{n/2}/({\ell |{\bf p}|})$
as  ${\bf p}\rightarrow\infty$, 
which shows that the divergence in $\Delta \epsilon_{\pm n}$ is 
of the form $\propto \pm \sqrt{n}\, \log (\ell\,  \Lambda)$, 
with momentum cutoff $\Lambda$.
Accordingly, for $M  \rightarrow 0$, 
the divergences in all $\Delta \epsilon_{n}$ are removed via renormalization of velocity, 
\begin{equation}
v = Z_{v}\, v^{\rm ren} = v^{\rm ren} + \delta v,
\end{equation}
with the counterterm $\delta v = (Z_{v}-1)\, v^{\rm ren}$.

Nonzero band gap $M=\omega_{c}\, \mu \not=0$ requires further renormalization~\cite{KS_CR}.  
It is not {\it a priori} clear how to renormalize $M$ for $B\not=0$.
A key step is to note that the magnetic field $B$ acts as a long-wavelength cutoff $\sim \ell$,
without affecting the short-distance structure of the theory.  
One can therefore determine the necessary counterterms from the $B=0$ theory, 
which yields
\begin{equation}
\delta v \sim -(\alpha/8 \epsilon_{b})\, \log \Lambda^{2},\
\delta M = 2\,  (M^{\rm ren}/v^{\rm ren})\, \delta v
\label{vctMct}
\end{equation}
(or equivalently, $\delta \mu = \mu^{\rm ren} \delta v/v^{\rm ren}$), 
with $M= M^{\rm ren} + \delta M$;
we denote (finite) renormalized quantities as 
$\omega_{c}^{\rm ren}\equiv \sqrt{2}\, v^{\rm ren}/\ell$, 
$M^{\rm ren}  =\omega_{c}^{\rm ren}\mu^{\rm ren}$ and 
$\epsilon_{n}^{\rm ren}= \omega_{c}^{\rm ren}\,e_{n}(\mu^{\rm ren})$.

 Rewriting the (bare) zeroth energy  as
$\epsilon_{n}=  \omega_{c}\, e_{n}= \epsilon_{n}^{\rm ren} + \delta_{\rm ct} \epsilon_{n}$
allows one to isolate the counterterm to $O(\alpha)$,
\begin{eqnarray}
\delta_{\rm ct} \epsilon_{n} &=& \omega_{c}^{\rm ren}\, \lambda_{n}(\mu^{\rm ren})\, 
\delta v/v^{\rm ren}, \nonumber\\
\lambda_{n}(\mu) &\equiv& e_{n} + \mu^2/e_{n} =  s_{n}(|n| + 2\mu^2)/ \sqrt{|n| + \mu^2}.
\label{ctEn_lambda}
\end{eqnarray}
Note that $\lambda_{n}(\mu) =s_{n} \sqrt{|n|} + O(\mu^2)$ for $n\not=0$ 
while $\lambda_{0_{-}}(\mu) =-2\mu$.
This implies that
$\Delta \epsilon_{n}$ are governed by a single divergence $\delta v$, 
which, for $n\not=0$, arises in all even powers of $\mu$.
In contrast, $\Delta \epsilon_{0}$ only has a divergence of $O(\mu)$. 
Direct calculation of $\sum_{\bf p} v_{\bf p}\, {\cal F}_{n}(z; \mu)$, indeed, verifies
such a nonlinear feature (in $M \sim \mu$) of renormalization; see Appendix A.

Let us denote the Coulomb-corrected level spectra as
\begin{equation}
\hat{\epsilon}_{n} = \epsilon_{n} + \Delta  \epsilon_{n}
= \epsilon_{n}^{\rm ren} + (\Delta  \epsilon_{n})^{\rm ren}.
\end{equation}
The renormalized self-energies 
$(\Delta  \epsilon_{n})^{\rm ren}\equiv \Delta  \epsilon_{n} + \delta_{\rm ct}\epsilon_{n}$
are now free of divergence.
One has to define the renormalized velocity $v^{\rm ren}$ 
by referring to some observable quantity.
Let us refer to $1\leftarrow 0_{-}$ resonance with zero band gap $\mu=0$ 
and let  
$\epsilon_{\rm  exc}^{1\leftarrow 0}|_{\mu=0} =\omega_{c}^{\rm ren}
\equiv \sqrt{2}\, v^{\rm ren}/\ell$
at each value of $B$; other choices are equally possible, as we remark later.
We thus choose 
$(\Delta \epsilon^{1, 0})^{\rm ren}|_{\mu=0}
\equiv (\Delta \epsilon^{1, 0} + \delta_{\rm ct} \epsilon_{1})|_{\mu=0} =0$,
 i.e., 
$\delta v = - (\ell/\sqrt{2})\, \Delta \epsilon^{1,0}|_{\mu=0}$ with
$\Delta \epsilon^{1,0}|_{\mu=0} = {1\over{2}} \sum_{\bf p}v_{\bf p}\, 
\big\{ {\cal F}_{1}(z;0) -e^{-z}\, (1- z/2) \big\}$.
One can then renormalize ${\cal F}_{n}(z;\mu)$ as
\begin{equation}
{\cal F}_{n}^{\rm ren}(z;\mu) 
=  {\cal F}_{n}(z;\mu) - \delta_{\rm ct}  {\cal F}_{n}(z;\mu),
\end{equation}
by isolating the portion that leads to a divergence,
\begin{equation}
\delta_{\rm ct}  {\cal F}_{n}(z;\mu) 
=\lambda_{n}(\mu)\,\{  {\cal F}_{1}(z;0) -e^{-z}\, (1- z/2) \},
\end{equation}
where ${\cal F}_{1}(z;0)  = e^{-z}\sum_{k=1}^{\infty}(k-z) z^{k-1}\sqrt{k}/k!$.
The self-energies 
$(\Delta \epsilon_{n})^{\rm ren}$ are thereby cast in a compact form
 \begin{eqnarray}
 (\Delta \epsilon_{n})^{\rm ren}  &=&\Omega_{n}(\mu)
- \sum_{k} \nu [k]\,  \sum_{\bf p}v_{\bf p} \gamma_{\bf p}^{2}  |g^{nk}_{\bf p}|^2, 
\label{DEn_ren} \\
 \Omega_{n}(\mu)&=&
{\textstyle{1\over{2}}}\sum_{\bf p}v_{\bf p}\, {\cal F}_{n}^{\rm ren}(z;\mu).
 \end{eqnarray}
In this renormalized form, 
${\cal F}_{n}^{\rm ren}(z;\mu)$ are sizable only for $\ell |{\bf p}| \sim O(1)$ 
and vanish rapidly as $z = {1\over{2}}\ell^{2}{\bf p}^{2} \rightarrow \infty$, 
and one can calculate $\Omega_{n}(\mu)$ numerically as well as analytically 
without handling divergences; 
see Appendix A for details. 
 $\Omega_{n}(\mu)$ enjoy the $eh$-conjugation property
 \begin{equation}
\Omega_{-n}(\mu) = - \Omega_{n}(-\mu),\ \ \Omega_{n}(\mu)|^{K'} = \Omega_{n}(-\mu)|^{K}.
\end{equation}

Similarly, the Coulomb-corrected CR energies are rewritten as 
$\epsilon_{\rm exc}^{n\leftarrow j} = \epsilon_{n}^{\rm ren}- \epsilon_{j}^{\rm ren} 
+ (\Delta \epsilon^{n,j})^{\rm ren}$,
with 
\begin{eqnarray}
(\Delta \epsilon^{n,j})^{\rm ren}
&=&    {\cal W}^{n,j}+  \sum_{k} \nu [k] \sum_{\bf p}v_{\bf p} \gamma_{\bf p}^{2}\,
\{ |g^{nk}_{\bf p}|^2 \! -\!  |g^{jk}_{\bf p}|^2\} ,
\nonumber\\
{\cal W}^{n,j} &=&    \Omega_{n} -  \Omega_{j} 
- (\nu_{j}\!-\nu_{n})\sum_{\bf p}v_{\bf p} \gamma_{\bf p}^{2}\, g^{nn}_{\bf -p}g^{jj}_{\bf p}, \ \ 
\label{DEnj_ren}
\end{eqnarray}
where $\Omega_{n}=\Omega_{n}(\mu)$ for short.
The renormalized corrections 
$(\Delta \epsilon_{n})^{\rm ren}$ and $(\Delta \epsilon^{n,j})^{\rm ren}$
are now divided into real-process contributions $(\propto \nu[k] )$ 
and many-body corrections $\Omega_{n}(\mu)$ and ${\cal W}^{n,j}$.

 \begin{table}
\caption{
Coulombic corrections $\Omega_{n}(\mu) \equiv \Omega_{n}$ 
and level shifts $(\Delta \epsilon_{n})^{\rm ren}|_{n_{\rm f}}$ (for $n=0,\pm 1)$
at valley $K$.
Setting $\mu\rightarrow-\mu$ yields 
$\Omega_{n}$ and$(\Delta \epsilon_{n})^{\rm ren}|_{n_{\rm f}}$ at valley $K'$. 
}
\begin{ruledtabular}
\begin{tabular}{| c | c |}
\multicolumn{2}{| l | } {\ $\Omega_{0_{-}}\! \approx \tilde{V}_{c}\,  (0.1688\, \mu)$} \\   
\hline
\ $\Omega_{1} \approx \tilde{V}_{c}\, ( 0.375\  + 0.0887\, \mu)$
\ & $\Omega_{4} \approx \tilde{V}_{c}\,  ( 0.3899 + 0.0288\, \mu) $\ \ \ \\
\ $ \Omega_{2} \approx \tilde{V}_{c}\, ( 0.4074 + 0.0520\, \mu)$ 
& $\Omega_{5}\approx \tilde{V}_{c}\,  (0.3673 + 0.0236\, \mu) $\ \ \  \\ 
\ $\Omega_{3} \approx \tilde{V}_{c}\,  (0.4054 + 0.0370\, \mu)$
& $\Omega_{6} \approx \tilde{V}_{c}\,  (0.3406 + 0.0200\, \mu) $\ \ \    \\
\end{tabular}
 \vskip0.1cm
\begin{tabular}{| l | c | c |c |}
\multicolumn{3}{| c } {$(\Delta \epsilon_{n})^{\rm ren}|_{n_{\rm f}} = \tilde{V}_{c}\, a_{n}|_{n_{\rm f}}$} 
&$n\in (-1,0_{-}, 1)$ \ \ \  \\   
\hline
$n_{\rm f}$ & -1&$0_{-}$ 
& 1 \\   
\hline
$a_{1}$ & ${1\over{2}} -0.0363\,\mu$ & ${1\over{4}}+0.2137\,\mu$ &  $ -{7\over{16}} + 0.0887\,\mu$\\
$a_{0_{-}}$ & ${1\over{2}} + 0.1688\,\mu$ &   $-{1\over{2}}+ 0.1688\,\mu$ & $-{3\over{4}} + 0.4188\,\mu$ \\ 
$a_{-1}$& $-{1\over{4}}+0.2137\,\mu$ & $- {1\over{2}} -0.0363\,\mu$ &   $ -{11\over{16}}  -0.0363\,\mu$  \\
\end{tabular}
\vskip0.1cm
\end{ruledtabular}
\end{table}



\begin{figure}[tbp]
\includegraphics[scale=0.55]{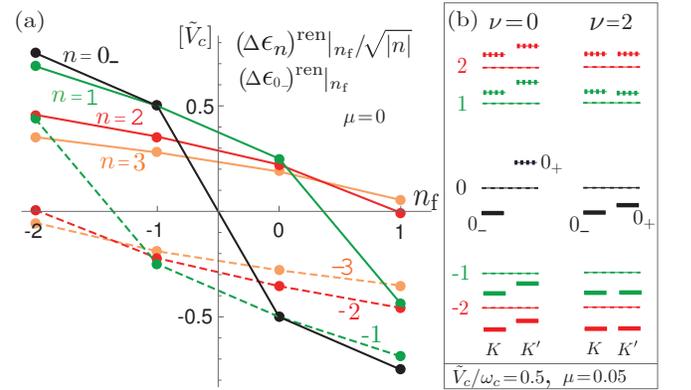}
\caption{(Color online) 
(a) Coulombic level shifts $(\Delta \epsilon_{0_{-}})^{\rm ren}|_{n_{\rm f}}$ and 
$(\Delta \epsilon_{n})^{\rm ren}|_{n_{\rm f}}/\sqrt{|n|}$ [in units of $\tilde{V}_{c}$]
for $\mu=0$ and  $n \in [-3,3]$ over the range $n_{\rm f} \in [-2,1]$.
Lines are a guide for the eyes.
(b)~An illustration of level spectra $\{ \hat{\epsilon}_{n} \}$ for $\nu=0$ and $\nu=2$, 
with $\tilde{V}_{c}/\omega_{c} \sim 0.5$ and $\mu\sim 0.05$ chosen tentatively.
}
\end{figure}


Table I shows a list of $\Omega_{n}(\mu)$ of our interest 
and $( \Delta \epsilon_{n})^{\rm ren}|_{n_{\rm f}}$ for $n=(-1,0_{-}, 1)$, 
 in units of 
\begin{equation}
\tilde{V}_{c} = {\alpha\over{\epsilon_{b}\, \ell}}\,  \sqrt{{\pi\over{2}}}
\approx {70.3\over{\epsilon_{b}}} \, \sqrt{B({\rm T})}\ {\rm meV};
\end{equation}
we suppose $\mu\ll 1$ and only retain terms to $O(\mu)$ below.
In Fig.~1(a), we depict level shifts 
$(\Delta \epsilon_{n\not=0})^{\rm ren}|_{n_{\rm f}}/\sqrt{|n|}$,
normalized relative to 
$\epsilon_{n \not=0}^{\rm ren}/ \sqrt{|n|} = s_{n}\, \omega_{c}^{\rm ren}$,
for $\mu=0$ and for a wider range $n \in [-3,3]$ and $n_{\rm f} \in [-2,1]$.
These shifts $(\Delta \epsilon_{n})^{\rm ren}$ 
 are generally sizable and, in particular,  $(\Delta \epsilon_{0})^{\rm ren}$ and 
 $(\Delta \epsilon_{\pm 1})^{\rm ren}$ critically change in magnitude 
 as the relevant $(0,\pm1)$ level is filled or emptied. 

Figure~1(b) illustrates a typical pattern of level spectra $\{ \hat{\epsilon}_{n} \}$ 
at the two valleys for the $\nu=0$ neutral state
and the $\nu=2$ state (with spin splitting suppressed).
[Note here that total filling $\nu=0$ refers to the valley content 
$(n_{\rm f}^{K}, n_{\rm f}^{K'})   = (0,-1)$,  
$\nu=-2$ to $(-1,-1)$, $\nu=2$ to $(0,0)$,
 $\nu=6$ to $(1,1)$, etc.]
For $\mu\not = 0$, Coulomb interactions lift the valley degeneracy of 
$n\not=0$ levels.
Valley asymmetry is minimum for $\nu= 4m+2 = \pm 2, \pm 6, \dots$, 
i.e., when large Landau gaps are present,
with $\hat{\epsilon}_{n \not =0}$ split only slightly $\sim O(\tilde{V}_{c}\,\mu)$ in the valley; 
see Eq.~(\ref{KKp_viamu}).
At $\nu=0$, in contrast, valley asymmetry is most prominent, 
especially for the $n=0$ levels, 
$( \Delta \epsilon_{0_{+}})^{\rm ren} -( \Delta \epsilon_{0_{-}})^{\rm ren} \sim \tilde{V}_{c}$.
This implies that even a tiny valley asymmetry can trigger a sizable Coulombic gap 
for the $\nu=0$ neutral state.

To survey CR or $\{T_{n}\}$, it is useful to handle excitation energies normalized as
$\epsilon^{n\leftarrow j}_{\rm exc}/N^{n,j} 
= \omega_{c}^{\rm ren} + (\Delta \epsilon^{n,j})^{\rm ren}/N^{n,j}$
with $N^{n,j} = s_{n}\sqrt{|n|} - s_{j}\sqrt{|j|} = N^{-j,-n}$;  $N^{1,0}=1$,
$N^{2,-1}= \sqrt{2}+1$, etc.
Table II shows a list of many-body corrections ${\cal W}^{n,-j}/N^{n,-j}$.
It is clear that Coulombic attraction dominates for intraband transitions
 while self-energy terms $\Omega_{n} -\Omega_{-j}$ dominate for interband transitions.
Table~II also summarizes the Coulombic corrections for $T_{2}$; 
similar tables for $T_{3}$ through $T_{6}$ are relegated to  Appendix A.
In Fig.~2, we depict all such corrections for $\mu=0$ and $n_{\rm f} \in [-3,2]$.
Clearly, for each $T_{n}$, peak values of 
$(\Delta \epsilon^{n,j})^{\rm ren}/N^{n,j}$ are associated with 
either $n_{\rm f}^{K}=0$ or  $n_{\rm f}^{K'} =-1$, i.e., with the $\nu=0$ state.

As seen from the scales of Figs.~1 and 2,
Coulombic contributions to CR are considerably smaller 
than level shifts $(\Delta \epsilon_{n})^{\rm ren}$; 
$(\Delta \epsilon^{n,-j})^{\rm ren}/N^{n,-j}$ 
are about 10 \% of $ \tilde{V}_{c}$ or less in magnitude for $T_{2} \sim T_{6}$
over the range $\nu \in [-2,2]$ (or  $n_{\rm f} \in [-1,0]$).
As for $T_{1}$, 
while $(\Delta \epsilon_{0_{\mp}})^{\rm ren}$ change abruptly
as one goes from $\nu=-2$ to $\nu=2$,
CR shifts $(\Delta \epsilon^{1,0})^{\rm ren}$ and $(\Delta \epsilon^{0,-1})^{\rm ren}$ 
remain far small, 
\begin{eqnarray}
(\Delta \epsilon^{1,0_{-}})^{\rm ren}
&=& (\Delta \epsilon^{0_{-}, -1})^{\rm ren}|_{\mu\rightarrow-\mu}
=  (\Delta \epsilon^{0_{+}, -1})^{\rm ren}, \nonumber\\
&\approx&
\tilde{V}_{c}\, \{ -0.2052\,\mu + 0.297\, \mu^2 + \cdots \};
\label{DEone-zero-r}
\end{eqnarray}
see Appendix A for details.


\begin{table}
\caption{ 
(a) Coulombic corrections ${\cal W}^{n,j}$ and 
(b) resonance shifts for $T_{2}$
(at valley $K$) in units of $\tilde{V}_{c}$; 
$N^{2,-1}= N^{1,-2} = \sqrt{2}+1$.
Setting $\mu\rightarrow-\mu$ yields those at valley $K'$. 
}
\begin{ruledtabular}
\vskip0.1cm
\begin{tabular}{| l | l |}
\multicolumn{2}{| l | } {\ \ \ ${\cal W}^{n,j}/N^{n,j} = \tilde{V}_{c}\, w^{n,j}$} \\   
\hline
$w^{2,1}\approx  -1.1666 - 0.0866\, \mu $ &  $w^{3,2}\approx -1.4380 - 0.0564\, \mu $\\
\hline
$w^{1,0_{-}}\approx  \textstyle -3/8 - 0.3302\, \mu $ &  $w^{4,-3}\approx 0.1016 - 0.0203\, \mu $\\
$w^{2,-1}\approx  0.1105 - 0.0796\, \mu $&  $w^{5,-4}\approx 0.0872 - 0.0137\, \mu $  \\ 
$w^{3,-2}\approx 0.1137 - 0.0347 \, \mu $&  $w^{6,-5}\approx 0.0729 - 0.0100 \, \mu $    \\
\end{tabular}
\vskip0.1cm
\begin{tabular}{| c | l | l |}
$n_{\rm f}$ &  $(\Delta \epsilon^{2,-1})^{\rm ren}|_{n_{\rm f}} /N^{2,-1}$ \   [$\tilde{V}_{c}$] 
& $(\Delta \epsilon^{1,-2})^{\rm ren}|_{n_{\rm f}} /N^{1,-2}$  \\   
\hline
-2 &  & 0.06742 + 0.0114  $\mu$ \\
-1 & 0.09756 - 0.159  $\mu$  & 0.1234\ \   + 0.00034 $\mu$\ \ \ \  \\ 
0 & 0.1234\ \  - 0.00034 $\mu$  & 0.09756 + 0.159  $\mu$ \\
1 & 0.06742 - 0.0114  $\mu$  & \\
\end{tabular}
\end{ruledtabular}
\end{table}



\begin{figure}[tbp]
\includegraphics[scale=0.6]{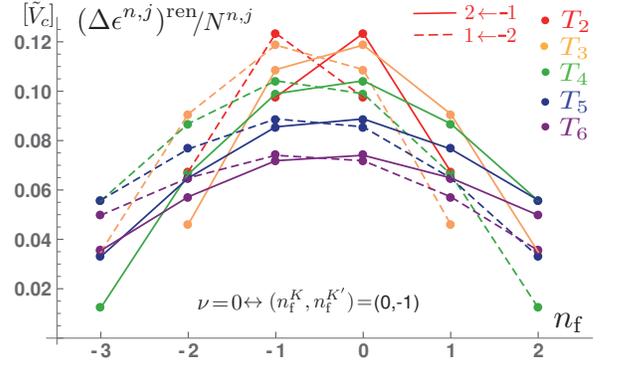} 
\caption{(Color online)  
Coulombic shifts $(\Delta \epsilon^{n,j})^{\rm ren}$  of CR for $\mu=0$.
Points guided by solid lines refer to the $n \leftarrow -(n-1)$ channel 
and those by dashed ones to the conjugate channel $n-1 \leftarrow -n$ of each $T_{n}$. 
}
\end{figure}


\section{Cyclotron resonance}

\begin{figure}[tb]
\includegraphics[scale=0.8]{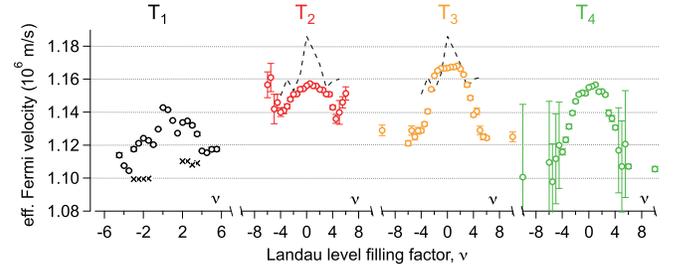} 
\caption{(Color online) 
$v^{\rm eff}$  vs $\nu$.  
A portion of experimental results reproduced from Ref.~\cite{RZTW}.
}
\end{figure}


Before proceeding further let us take a quick look at some recent experimental data of Ref.~\cite{RZTW}.
There the observed CR energies are parametrized in the one-body form
$\epsilon^{n\leftarrow j}_{\rm exc}|^{\rm exp} \equiv (\sqrt{2}\,   v^{\rm eff} /\ell)\, N^{n,j}$,
and the effective Fermi velocity $v^{\rm eff}$ is determined 
for six major transitions $T_{1}$ through $T_{6}$.
Figure~3 reproduces a portion of the data, in which 
$v^{\rm eff}|^{\rm exp}$ at $B=8$T is plotted as a function of filling factor $\nu$
for $T_{1} \sim T_{4}$.
$v^{\rm eff}|^{\rm exp}$ refers to the whole active channels of each $T_{n}$ 
and varies by $2\% \sim 5 $\% over the range $|\nu| \lesssim 10$.
At a glance, for all $T_{1} \sim T_{6}$, variations of $v^{\rm eff}$ with $\nu$
are nearly symmetric about $\nu=0$ with a maximum at $\nu=0$.
This provides, as noted regarding Eq.~(\ref{Eex-conj}), 
direct evidence that $eh$ conjugation is well realized in graphene.
For $T_{3}$ $\sim T_{6}$, $v^{\rm eff}$ shows a generally similar $\nu$ dependence.
In contrast, for $T_{1}$, a splitting of $v^{\rm eff}$ is seen around $\nu \sim \pm 2$,
and, for $T_{2}$, $v^{\rm eff}$ shows minima around $\nu \sim \pm 4$.
With such data in mind, let us continue our analysis.

$T_{1}$ is rather special from the viewpoint of the SMA: 
As noted in Eq.~(\ref{DEone-zero}) or in Eq.~(\ref{DEone-zero-r}), 
Coulombic corrections $\Delta \epsilon^{1,0}$ and $\Delta \epsilon^{0,-1}$ 
are independent of $\nu_{0_{\pm }}$ 
(i.e., filling of $n=0_{\pm}$ levels) and are entirely due to many-body effects. 
 In general, CR, being spin preserving, is unaffected by spin splitting 
 (which actually  is rather small for $B <10$T).
It appears difficult to interpret the observed $T_{1}$ data by Coulombic contributions alone.
This naturally leads us to a possible band gap $M$.


\begin{figure}[tbp]
\includegraphics[scale=0.85]{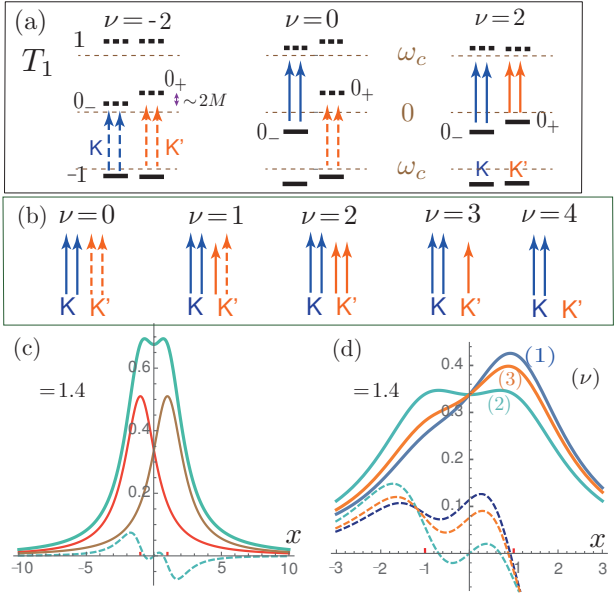} 
\caption{(Color online) 
$T_{1}$  (a)  Active resonance channels 
of $T_{1}$ vary with total filling $\nu \in (-2,0,2)$,
with $\tilde{V}_{c}/\omega_{c}\sim 0.5$ and $\mu\sim 0.1$ chosen tentatively; 
spin splitting is suppressed. 
(b) Composition of active resonances for $\nu =0\sim 4$.
Split spectra arise for $\nu \not=(0,4)$.
(c) An apparent reduction of splitting via broadening.
In the superposition of two Lorentzians $f_{\pm} (x)=1/\{w^2 + (x\mp d)^2 \}$
with $(d, w)=(1, 1.4)$, splitting ($2d$) of peaks is apparently reduced by $\sim $30\%.
The dashed line stands for a derivative of the total profile. 
(d) Simulated resonance profiles for the $\nu = (1,2,3)$ configurations in (b), 
with basic profiles $f_{\pm} (x)$ used.  
The peak position $x=1$ is apparently shifted to $x\approx (0.92, 0.69, 0.88)$ 
for $\nu=(1,2,3)$, respectively.
}
\end{figure}


Figure~4(a) is an illustration of  the level spectra and associated $T_{1}$ spectra
for $\nu=(-2, 0, 2)$. 
[Note here $eh$ conjugation: 
The $\nu=0$ level spectra 
remain the same under level inversion $n\rightarrow-n$ about $n=0$ and 
$K\leftrightarrow K'$, while the $\nu=-2$ level spectra thereby
turn into the $\nu=2$ level spectra.]
The range $\nu \in [-2,2]$ concerns filling of $n=0_{\pm}$ levels 
and, over this range,
$T_{1}$ consists of  $1 \leftarrow 0$ and  $0 \leftarrow -1$ transitions,
with resonance energy 
\begin{eqnarray}
\epsilon^{1\leftarrow 0_{-}}_{\rm exc} &=& \epsilon^{0_{+} \leftarrow-1}_{\rm exc}  
= \omega_{c} \{1+ (1- \xi)\mu  +\cdots \},
\nonumber\\
\epsilon^{0_{-} \leftarrow -1}_{\rm exc}  &=&\epsilon^{1 \leftarrow 0_{+}}_{\rm exc} 
=   \omega_{c} \{1 - (1- \xi)\mu + \cdots \},
\\
\xi &\approx&  0.205\, \tilde{V}_{c}/\omega_{c},
\end{eqnarray}
where the $O(\xi \mu)$ terms come from $(\Delta \epsilon^{1,0_{-}})^{\rm ren}$
in Eq.~(\ref{DEone-zero-r}).
Here and from now on, $(\omega_{c}, \mu, M, \cdots)$ refer to renormalized quantities.

Figure~4(b) illustrates how active resonance channels change in content 
as $\nu$ is increased from $0$ to $4$ under fixed $B$. 
At $\nu=0$ the four (valley, spin) channels have the same resonance energy 
$\omega_{c} + (1- \xi)\, M$.
At $\nu=1$ one of the spin-split $0_{+}$ levels gets filled and 
a $1 \leftarrow 0_{+}$ channel becomes active in place of $0_{+} \leftarrow -1$.
At $\nu=2$ the resonance energies are split in the valley by $2(1- \xi)\,M$.
At the same time, the $n=1$ levels are slightly split in the valley, with
 valley $K'$ lower in energy by $\sim0.43\, \tilde{V}_{c}\, \mu$; 
 see $ (\Delta \epsilon_{1})^{\rm ren}|_{n_{\rm f}=0}$ in Table~I. 
Accordingly, it is valley $K'$ that is first filled as one goes to $\nu = 3$.
At $\nu=4$ only the $1\leftarrow 0_{-}$ channels  remain active. 

In practice, such resonances
are broadened under disorder and their profiles overlap each other.
With increasing disorder, a splitting of competing resonances will become less prominent 
and, when splitting and broadening are comparable, 
resonance peaks will start to shift in position to eventually merge into a single broad profile; 
one would thereby observe an apparent reduction in splitting and in resonance energy, 
as illustrated in Fig.~4(c).

The resonance spectra in Fig.~4(b) change in both number and energy with $\nu$.
To simulate the effect of disorder, let us consider, at each $\nu$, 
an average of competing resonances, with a certain spread added, 
and determine the peak positions of the spectra.
Figure 4(d) shows such simulated resonance profiles 
for the $\nu=(1,2,3)$ configurations in Fig.~4(b),
in which an apparent reduction in peak energy and in splitting is seen.
A peak of $v^{\rm eff}$ at $\nu=0$, its splitting at $\nu\sim \pm 2$ and 
a subsequent partial rise of $v^{\rm eff}$ for $|\nu| \gtrsim 2$ 
in the $T_{1}$ data are qualitatively consistent with this picture of theory.
Note, in this connection, a crucial effect of small band gap $M$: 
If, at $\nu=2$, the $n=1$ level were split in the valley so that valley $K$ is lower, 
 one would observe a further decrease of 
 $v^{\rm eff}$ in going from $\nu=2$ to $\nu=4$.

\section{Interband resonance $T_{2}$ $\sim$ $T_{6}$ }


\begin{figure}[tbp]
\includegraphics[scale=0.65]{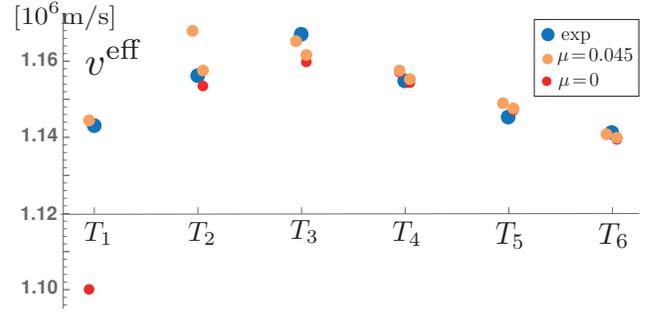} 
\caption{(Color online)   
Experimental peak values (blue points) of $v^{\rm eff}$ at $\nu=0$ and $B=8$T,
plotted for $T_{1} \sim T_{6}$.
Also plotted are theoretical values of  $v^{\rm eff}$ for $T_{n}$, 
with the choice $v^{\rm ren} = 1.1\times 10^{6}$m/s and 
$\tilde{V}_{c}/\omega_{c} = 0.5$;
orange points refer to the choice $\mu= M/\omega_{c}=0.045$ and 
red points to zero gap $\mu=0$. 
Points are slightly displaced horizontally to distinguish conjugate channels. 
}
\end{figure}


Figure~5 shows (by blue points) a plot of $v^{\rm eff}$ at $\nu=0$  and $B=8\, {\rm T}$ 
extracted from the $T_{1} \sim T_{6}$ data of Ref.~\cite{RZTW}.
These peak values of $v^{\rm eff}|^{\rm exp}$ rise as one goes from $T_{1}$ to $T_{3}$ 
and then decrease for higher $T_{n}$.
Also included is a plot of 
$v^{\rm eff}|^{\rm theory}\propto \epsilon^{m\leftarrow j}_{\rm exc}/ N^{m,j}$ 
at $\nu=0$ for each $T_{n}$, 
with the $n \leftarrow -(n-1)$ and 
$n-1 \leftarrow -n$ channels distinguished, and 
with the velocity chosen to be 
\begin{equation}
v^{\rm ren} = 1.1 \times 10^{6}\,  {\rm m/s}.
\label{vren_choice}
\end{equation}

Theoretical values $v^{\rm eff}|^{\rm theory}$ depend sensitively 
on the magnitude of $\tilde{V}_{c}$ and band gap $M$, and 
show a characteristic monotonous decrease in going from $T_{3}$ to $T_{6}$, 
as seen from Fig.~2. 
Adjusting the gradient gives 
\begin{equation}
\tilde{V}_{c}/\omega_{c} \approx 0.5 \sim 0.6.
\label{VoverWc}
\end{equation}
$T_{1}$ critically depends on band gap $M$.   Adjusting its position yields 
\begin{equation}
\mu \approx 0.04 \sim 0.05.
\label{mu_theor}
\end{equation}
In Fig.~5 orange points refer to the plot of $v^{\rm eff}|^{\rm theory}$ 
with the choice $\tilde{V}_{c}/\omega_{c} =0.5$ and $\mu=0.045$,
which gives a practically good fit
to the experimental data;
for comparison, red points refer to the case of zero band gap $\mu=0$. 
The choice of $v^{\rm ren}$ in Eq.~(\ref{vren_choice}) 
is also a practically unique choice 
since a slight change of it shifts the theoretical plot vertically and almost uniformly.

It will be worthwhile to remark here that relative variations of $v^{\rm eff}$ 
among  $T_{1} \sim T_{6}$ have definite meaning independent 
of  the choice of renormalization prescriptions.
Actually, upon adopting a new prescription, 
renormalized spectra $\epsilon_{n}^{\rm ren}$ and self-energies $\Omega_{n}(\mu)$ change 
but the the sums $\epsilon_{n}^{\rm ren}+\Omega_{n}(\mu)$ remain invariant, 
as explained in Appendix B.
Thus the best-fit values of $(v^{\rm ren}, \mu^{\rm ren}, \cdots)$ here are simply translated 
to another equivalent set in the new prescription,  with no change in physics.

The above choice of $(v^{\rm ren}, \tilde{V}_{c}/\omega_{c}, \mu)$ in turn leads to a band gap
\begin{equation}
2M = 2\, \omega_{c} \, \mu \sim 10\, {\rm meV}
\end{equation}
and $\tilde{V}_{c} \sim 60\, $meV or  $\epsilon_{b} \sim 3$ at $B= 8$T.  
A band gap of $\sim 10\,$meV amounts to 
an $\sim 8\, \%$ splitting of $v^{\rm eff}$ at $\nu\sim \pm 2$ for $T_{1}$,
which appears considerably larger than the observed $\sim 3$\% splitting of $v^{\rm eff}$ in Fig.~3. 
This will nevertheless be a reasonable estimate in view of an apparent reduction 
of a splitting in the presence of disorder, illustrated in Fig.~4(d).

Actually, some earlier experiments reported observations of larger band gaps of $\sim $30 meV 
in graphene/hBN devices~\cite{HuntYY,CSYL,WoodsBE} 
and the possibility of much smaller gaps in encapsulated devices~\cite{WoodsBE},
as also discussed theoretically~\cite{GKBK,SGS,JDMA}.  
The gap of $\sim $10 meV in an encapsulated device here is in accord with the latter.

Let us next examine $T_{2}$ in more detail.
It is enlightening to first  look at a short summary of  corrections 
$\{ (\Delta \epsilon^{2,-1})^{\rm ren}/N^{2,-1},  
(\Delta \epsilon^{1,-2})^{\rm ren}/N^{1,-2}\}_{n_{\rm f}}$
[in units of $\tilde{V}_{c}$]
at each valley (per spin) for $\nu \in (0,2,4,6)$:
\begin{eqnarray}
&&0: \{0.123_{-}, 0.097_{+}\}^{K}_{n_{\rm f}=0}, \{  0.097_{+}, 0.123_{-}\}^{K'}_{n_{\rm f}=-1},
\nonumber\\
&&2: \{0.123_{-}, 0.097_{+}\}^{K}_{n_{\rm f}=0}, \{  0.123_{+}, 0.097_{-}\}^{K'}_{n_{\rm f}=0},
\nonumber\\
&&4: \{0.123_{-}, 0.097_{+}\}^{K}_{n_{\rm f}=0}, \{  0.067_{+}, \ \ \  -  \ \ \ \}^{K'}_{n_{\rm f}=1},
\nonumber\\
&&6: \{0.067_{-}, \ \ \  -  \ \ \ \}^{K}_{n_{\rm f}=1}, \{  0.067_{+}, \ \ \  -  \ \ \ \}^{K'}_{n_{\rm f}=1},
\label{Ttwo_summary}
\end{eqnarray}
where $_{\pm}$ refers to the sign of $O(\mu)$ corrections, 
e.g., $0.097_{\pm } \equiv 0.97 \pm 0.16 \mu \approx 0.105/0.09$ (with $\mu \sim 0.05)$;
$0.123_{\pm} \approx 0.123$ and $0.067_{\pm } \approx 0.068/0.067$.
At $\nu=0$, the excitation spectra, though split to $\sim 2\%$ of  $\tilde{V}_{c}$, 
are the same at both valleys 
while at $\nu=2$ they are further split  in the valley to $O(\mu \tilde{V}_{c})$;
such weak splitting, unlike in $T_{1}$, may well be invisible under disorder.
In this way, $v^{\rm eff}$ will have a maximum at $\nu=0$ and 
barely change over the range $-2 \le \nu \le 2$.
This feature of $v^{\rm eff}$ is common to other $T_{n}$ as well, 
and is consistent with the experimental data on $T_{2} \sim T_{6}$.

As one goes from $\nu=2$ to $\nu=4$, the $n=1$ levels at valley $K'$ are gradually filled, 
with the $(1\leftarrow -2)^{K'}$ channel closed and $(\Delta \epsilon^{2,-1})^{\rm ren}|^{K'}$
reduced in magnitude.  One will therefore observe a decrease of $v^{\rm eff}$ 
in going to $\nu=4$ and, according to Eq.~(\ref{Ttwo_summary}), even further to $\nu=6$.
The present theory is thus consistent with the observed variation of $v^{\rm eff}$ 
over the range $-4 \le \nu \le 4$ in the $T_{2}$ data.
The unexpected rise of  $v^{\rm eff}$ from $\nu= \pm 4$ to $\nu= \pm 6$ in the $T_{2}$ data, 
however, remains unexplained~\cite{fn_suppl}.


\begin{figure}[tbp]
\includegraphics[scale=0.58]{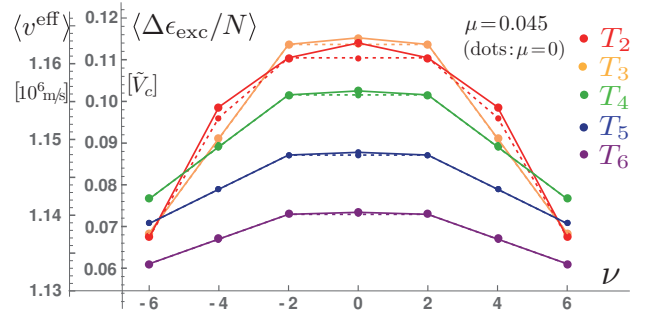} 
\caption{(Color online)
Coulombic corrections $\Delta \epsilon_{\rm exc}^{n,-j}/N^{n,-j}$ 
averaged over active channels of each $T_{n}$ for $|\nu|\le 6$;
the scale is also translated to averaged velocity $\langle v^{\rm eff} \rangle$ 
using $v^{\rm ren} = 1.1 \times 10^{6}{\rm m/s}$ and 
$\tilde{V}_{c}/\omega_{c}=0.5$.
Points guided by real lines refer to the case $\mu=0.05$ 
and those by dotted lines to zero band gap $\mu=0$.
}
\end{figure}


As in the observed $T_{2}\sim T_{6}$ data, when competing resonances merge 
into a single broad peak in the presence of disorder, 
its peak position will lie around the average of the resonance spectra.
To simulate such $\nu$ variations of $v^{\rm eff}$ 
we show in Fig.~6 
the Coulombic corrections $\Delta \epsilon_{\rm exc}^{n,-j}/N^{n,-j}$ 
averaged over (eight or less) active channels of each $T_{n}$ 
for $-6\le \nu \le 6$;
the scale is also translated to averaged velocity $\langle v^{\rm eff} \rangle$,
using $v^{\rm ren} = 1.1 \times 10^{6}{\rm m/s}$,
$\tilde{V}_{c}/\omega_{c}=0.5$ and $\mu=0.05$.
This figure compares well with the observed data in Fig.~3, and 
favorably explains why $v^{\rm eff}$ attains the largest peak value for  $T_{3}$ 
rather than $T_{2}$ at $\nu=0$
while, as emphasized in Ref.~\cite{RZTW}, 
the peak value shifts to $T_{2}$ at $\nu=\pm4$.
Quantitatively, however, the decrease of $\langle v^{\rm eff} \rangle$ 
over the range $2\le |\nu| \le 6$ for $T_{3} \sim T_{6}$ 
is twice or more slower than the observed 2\% - 4\% decrease of $v^{\rm eff}$.
Presumably this discrepancy is attributed to the presence of screening in $v_{\bf p}$.
For graphene the effect of screening grows with increasing $|\nu|$ 
more rapidly~\cite{KS_screening,SZL} than for GaAs heterostructures. 
 The Coulomb potential $v_{\bf p}$ will thus get weaker with increasing $|\nu|$, 
 making the decrease of  $v^{\rm eff}$ steeper for $|\nu| >2$ in Fig.~6.

 \section{Summary and discussion}
 
 In this paper 
we have studied Coulombic corrections to Landau-level spectra and CR in graphene, 
with a band gap taken into account.
Theory based on the SMA turns out to well explain, at least qualitatively,  
the recent experimental data of Ref.~\cite{RZTW}, which measured, in particular,
how the effective velocity $v^{\rm eff}$ varies as a function of filling factor $\nu$ 
under fixed field $B$
for six sets $T_{1} \sim T_{6}$ of leading interband CR.
Many-body corrections to level spectra, 
though not directly detectable, are generally sizable 
while those to observable CR signals turn out to be
much smaller, about 10\% of $\tilde{V}_{c}$ or less in magnitude.

The presence of Coulombic effects is clearly seen from relative variations 
of the peak values of $v^{\rm eff}$ (at $\nu=0$) among 
$T_{3}$ through $T_{6}$.
The presence of a small band gap 
 is inferred from a unique variation of $v^{\rm eff}$ 
in the $T_{1}$ data, in which 
the $1\leftarrow 0$ and $0\leftarrow -1$ resonance channels compete.  
The data suggest a band gap of $2M \sim 10$ meV at $B=8$T.

Particular attention has been paid to $eh$ conjugation symmetry, 
that relates the level and CR spectra at the two valleys.
Each set $T_{n}$ consists of an $eh$-conjugate pair 
[$n \leftarrow -(n-1)$ and $n-1 \leftarrow -n$ ] 
of CR channels, which differ slightly by Coulombic corrections.
The observable signal $v^{\rm eff}$, 
under $eh$ conjugation,
has a profile symmetric in $\nu$ with a maximum at $\nu=0$ for each $T_{n}$, 
which is indeed seen as a notable feature in the observed data.

Interband CR, specific to graphene and observable over a certain continuous range of filling factor 
$\nu$, is a useful window to explore many-body effects.
It is highly desired that experiment in this direction be extended to 
few-layer graphene and related Dirac-electron systems, 
where interesting many-body and topological quantum phenomena come into play.


\appendix 

\section{Coulombic corrections}

In this Appendix we outline calculations of the Coulombic corrections $(\Delta \epsilon_{n})^{\rm ren}$ 
and $(\Delta \epsilon^{n,j})^{\rm ren}$ in Eqs.~(\ref{DEn_ren}) and (\ref{DEnj_ren}).  
For ${\cal F}_{n}(z;\mu)$
it is useful to rewrite $|g^{nk}_{\bf p}|^2 - |g^{n,-k}_{\bf p}|^2\equiv \Gamma/(e_{n}e_{k})$, with
\begin{eqnarray}
\Gamma &=& {\mu\over{2}}\,(e_{n}- \mu) \{   |f^{N-1,K-1}_{\bf p}|^2 - |f^{N,K}_{\bf p}|^2\}  
\nonumber\\
&&+ \mu^2\,  |f^{N-1,K-1}_{\bf p}|^2 
+ \sqrt{NK}\, X^{NK}_{\bf p},
\end{eqnarray}
where $N\equiv |n|$, $K\equiv |k|$ and $X^{NK}_{\bf p}= {\rm Re}[ f^{N-1,K-1}_{\bf p}f^{K,N}_{\bf -p} ]$. 
The finite  $O(\mu)$ term in ${\cal F}_{n}(z;\mu)$ is uniquely determined from the first term 
while the remaining terms contain the ultraviolet divergence proportional to $(e_{n}^2 + \mu^2)/e_{n}$.

An efficient way to calculate 
$2\, \Omega_{n}(\mu)=\sum_{\bf p}v_{\bf p}\, {\cal F}_{n}^{\rm ren}(z;\mu)$ 
numerically is to first evaluate ${\cal F}_{n}^{\rm ren}(z;\mu)$ 
by summing sufficiently many terms in it
so that subsequent integration over ${\bf p}$ is dominated 
by the small-momentum domain 
$\ell |{\bf p}|\lesssim 1$.
An alternative way, suited for analytic treatment, is to integrate over ${\bf p}$ first,
$\sum_{\bf p} v_{\bf p} f =\tilde{V}_{c} \int_{0}^{\infty}(dz/\sqrt{z \pi})\,f$.
This yields, e.g., 
\begin{eqnarray}
\sum_{\bf p}v_{\bf p}\, {\cal F}_{1}(z;\mu)
&=& \tilde{V}_{c} \left[  
{1+ 2\mu^2\over{e_{1}}}\,{\cal D}+  {\mu\over{4}}\, (1-  {\mu\over{e_{1}}}) \,  \Xi 
\right], \ \ \ 
\\
({\cal D}, \Xi) &=&
{1\over{2\sqrt{\pi}}} \sum_{k=1}^{\infty}{(k, 1)\over{e_{k}}}\, {1\over{k!}}\, \Gamma (k -1/2).
\end{eqnarray}
One can write 
\begin{eqnarray}
{\cal D} &=& {\cal D}_{0}- \mu^2 \Xi_{0}/2 + O(\mu^4), 
\nonumber\\
\Xi &=& \Xi_{0} + O(\mu^2),\ \  \Xi_{0} \approx 0.7092,
\end{eqnarray}
where $({\cal D}_{0}, \Xi_{0})\equiv ({\cal D}, \Xi)|_{\mu\rightarrow0}$ (i.e., $e_{k}\rightarrow \sqrt{k}$).
Only ${\cal D}_{0}$ has a logarithmic divergence $\propto \ln k_{\rm cutoff}$.
Isolating the counterterm $\sum_{\bf p}v_{\bf p}\delta_{\rm ct}  {\cal F}_{n}(z;\mu)
= \lambda_{n}(\mu)\,\tilde{V}_{c} ({\cal D}_{0} - 3/4)$, 
one can calculate 
$\Omega_{n}(\mu)=
{1\over{2}}\sum_{\bf p}v_{\bf p}\, {\cal F}_{n}^{\rm ren}(z;\mu)$,
e.g., 
\begin{equation}
\Omega_{1}(\mu) = \textstyle
   \tilde{V}_{c}\, \Big[ {3\over{8}} +{1\over{8}}\, \Xi_{0}\,\mu 
   +( {9\over{16}} -  {3\over{8}}\Xi_{0} )\, \mu^2 +O(\mu^3) \Big].
\end{equation}
We have checked by direct calculations that both ways of calculation lead to the same result for 
all $\Omega_{n\not=0}(\mu)$ listed in Table I.


 \begin{table}
\caption{
Coulombic CR shifts $(\Delta \epsilon^{n,-(n-1)})^{\rm ren}|_{n_{\rm f}}$ (at valley $K$)
in units of $\tilde{V}_{c}$ for $T_{3} \sim T_{6}$.
Setting $\mu\rightarrow-\mu$ yields those at valley $K'$,
while reversing $(\Delta \epsilon^{n\leftarrow -(n-1)})^{\rm ren}|_{n_{\rm f}}$
in $n_{\rm f}$ about $n_{\rm f}= -1/2$ 
yields $(\Delta \epsilon^{n-1\leftarrow -n})^{\rm ren}|^{K'}_{-n_{\rm f}-1}$. 
}
\begin{ruledtabular}
\vskip0.1cm
\begin{tabular}{| c | l | l |}
$n_{\rm f}$ &  $(\Delta \epsilon^{3,-2})^{\rm ren}/N^{3,-2}$  & $(\Delta \epsilon^{4,-3})^{\rm ren}/N^{4,-3}$  \\   
\hline
-2 & 0.04606 - 0.0627 $\mu$  & 0.06643 - 0.0389 $\mu$ \  \ \\
-1 & 0.1087 - 0.0701  $\mu$  &0.09898 - 0.0416  $\mu$ \\ 
0 & 0.1187 + 0.000731  $\mu$\ \ \ \   &0.1042\ \  + 0.000933  $\mu$\ \ \ \  \\
1 & 0.09076 - 0.00173  $\mu$  & 0.08684 - 0.000144  $\mu$\\
2 & 0.03489 - 0.00572 $\mu$  & 0.05562 - 0.00162 $\mu$\\
\end{tabular}
\vskip0.1cm
\begin{tabular}{| c | l | l |}
$n_{\rm f}$ &  $\Delta \epsilon^{5,-4}/N^{5,-4}$  &$\Delta \epsilon^{6,-5}/N^{6,-5}$  \\   
\hline
-2 & 0.06485 - 0.0270$\mu$ & 0.05722 - 0.0200 $\mu$\\
-1 &  0.08555 - 0.0282 $\mu$  & 0.07185 - 0.0208  $\mu$ \\ 
0 &0.08878 + 0.000886  $\mu$  & 0.07404 + 0.000790  $\mu$ \\
1 &0.07661 + 0.000295   $\mu$  & 0.06489 + 0.000423  $\mu$\\
2 & 0.05587 - 0.000462 $\mu$  & 0.04979 - 0.0000281 $\mu$\\
\end{tabular}
\end{ruledtabular}
\end{table}


Some care, on the other hand, is needed for $\Omega_{0_{-}}(\mu)$, 
which is written as 
$\Omega_{0_{-}}(\mu)= \mu\tilde{V}_{c} \{- 3/4 + \Xi^{\rm reg}/2\}$ with
\begin{eqnarray}
\Xi^{\rm reg} &=&{ 1\over{\sqrt{ \pi} }} \int_{0}^{\infty}{dz\,  e^{-z}\over{\sqrt{z}}}\, 
\sum_{k=1}^{\infty} {2k (k-z)z^{k-1} -z^{k}\,  \over{\sqrt{k}\ k!}},
\nonumber\\
&\approx& 1.8377.
\end{eqnarray}
This number is obtained by first summing over $k$ and then integrating over $z$,
which is a physically sensible step of calculation.
In contrast, if the step is reversed, one obtains $\Xi_{0} \approx 0.7092$. 
The difference presumably comes from a surface term.
Actually one can eliminate this $\Omega_{0_{-}}(\mu) \sim O(\mu \tilde{V}_{c})$ 
by an $O(\tilde{V}_{c})$ redefinition of $\mu^{\rm ren}$, 
without affecting $O(\tilde{V}_{c})$ corrections in all other $\Omega_{n}(\mu)$.

Finally we record $(\Delta \epsilon^{1,0_{-}})^{\rm ren}$,
\begin{eqnarray}
(\Delta \epsilon^{1,0_{-}})^{\rm ren}|^{K} &=&
\textstyle  \tilde{V}_{c}\, 
 \left\{  ({5\over{8}} + {1\over{8}}\, \Xi_{0}-{1\over{2}}\, \Xi^{\rm reg} )\,\mu 
+\cdots  \right\},
\nonumber\\
&\approx&
\tilde{V}_{c}\, \{ -0.2052\,\mu + 0.297\, \mu^2 + \cdots \},
\end{eqnarray}
which is quoted in Eq.~(\ref{DEone-zero-r}).
Table III records some principal portion of Coulombic contributions for $T_{3} \sim T_{6}$.

\section{Renormalization prescriptions}

In this appendix 
we examine how observable quantities depend on renormalization prescriptions adopted. 
So far we have used $v^{\rm ren}$ defined by referring to the $1\leftarrow 0_{-}$ resonance channel, 
with the counterterm 
$\delta v = - (\ell/\sqrt{2})\, \Delta \epsilon^{1,0}|_{\mu=0}$.
Suppose now that we refer to some other channel or other prescription by setting
$\delta v^{\rm new} = \delta v - (\ell/\sqrt{2})\, \Delta C$,
where the finite difference $\Delta C \propto O(\tilde{V}_{c})$ 
comes from the soft momentum domain $\ell |{\bf p}|\lesssim 1$.
One then passes, noting Eqs.~(\ref{vctMct}) and~(\ref{ctEn_lambda}), 
to the new renormalized quantities to $O(\tilde{V}_{c})$,
\begin{eqnarray}
v^{\rm ren;new} &=&  v^{\rm ren}+(\ell/\sqrt{2})\, \Delta C,
\nonumber\\
\mu^{\rm ren;new} &=& (1+ \Delta C/\omega_{c}^{\rm ren})\, \mu^{\rm ren},
\nonumber\\
M^{\rm ren;new} &=& (1+ 2\Delta C/\omega_{c}^{\rm ren})\, M^{\rm ren}.
\end{eqnarray}

At the same time, spectra $\epsilon_{n}^{\rm ren}$ and many-body corrections 
$\Omega_{n}(\mu)={1\over{2}} \sum_{\bf p}v_{\bf p}\, {\cal F}^{\rm ren}_{n}(z;\mu)$
get shifted, e.g., $\Omega^{\rm new}_{n}(\mu) 
= \Omega_{n}(\mu)  -\lambda_{n}(\mu)\,  \Delta C$, 
but the sum 
\begin{equation}
\epsilon_{n}^{\rm ren;new}+\Omega^{\rm new}_{n}(\mu)  
= \epsilon_{n}^{\rm ren} +\Omega_{n}(\mu)
\end{equation}
remains invariant  to $O(\tilde{V}_{c})$ under renormalization.  
In this way, one can transform the whole set of renormalized quantities and quantum corrections 
to another equivalent set in the new prescription.



\begin{thebibliography}{99}

\bibitem{NG} K.~S.~Novoselov, A.~K. Geim, S.~V.~Morozov, D.~Jiang,
 M.~I.~Katsnelson, I.~V.~Grigorieva, S.~V.~Dubonos, and 
A.~A.~Firsov, Nature (London) {\bf 438}, 197 (2005).

\bibitem{ZTSK} Y. Zhang, Y.-W. Tan, H. L. Stormer, and P. Kim, 
 Nature (London) {\bf 438}, 201 (2005).

\bibitem{GN_rev} A. K. Geim and K. S. Novoselov, Nat. Mater. {\bf 6}, 183 (2007).

\bibitem{AbF} 
D.~S.~L.~Abergel and V.~I.~Fal'ko, Phys. Rev. B {\bf 75}, 155430 (2007).

\bibitem{Kohn} W. Kohn, Phys. Rev. {\bf 123}, 1242 (1961).  

\bibitem{KH} C. Kallin and B. I. Halperin,  
Phys. Rev. B {\bf 30}, 5655 (1984).

\bibitem{GGV} J. Gonz\'alez, F. Guinea, and M.A.H. Vozmediano, 
Nucl. Phys. {\bf B} 424, 595 (1994).

\bibitem{IWF} A. Iyengar, J. Wang, H. A. Fertig, and L. Brey, Phys. Rev. B {\bf 75},
125430 (2007).

\bibitem{BM} Yu. A. Bychkov, and G. Martinez, 
Phys. Rev. B {\bf 77}, 125417 (2008).

\bibitem{KS_CR}  K. Shizuya, Phys. Rev. B {\bf 81}, 075407 (2010);
Phys. Rev. B {\bf 84}, 075409 (2011).

\bibitem{RFG} R. Rold\'an, J.-N. Fuchs, and M. O. Goerbig, 
Phys. Rev. B {\bf 82}, 205418 (2010).

\bibitem{JHTWS} Z. Jiang, E. A. Henriksen, L. C. Tung, Y.-J. Wang, M. E. Schwartz,
M. Y. Han, P. Kim, and H. L. Stormer, 
Phys. Rev. Lett. {\bf 98}, 197403 (2007).   

\bibitem{DCNN}R. S. Deacon, K.-C. Chuang, R. J. Nicholas, 
K. S. Novoselov, and A. K. Geim, 
Phys. Rev. B {\bf 76}, 081406(R) (2007).

\bibitem{HCJL} E. A. Henriksen, P. Cadden-Zimansky, Z. Jiang, Z. Q. Li, L.-C. Tung, 
M. E. Schwartz, M. Takita, Y.-J. Wang, P. Kim, and H. L. Stormer, 
Phys. Rev. Lett. {\bf 104}, 067404 (2010).

\bibitem{MNEB}  L. M. Malard, J. Nilsson, D. C. Elias, J. C. Brant, F. Plentz, 
E. S. Alves, A. H. Castro Neto, and M. A. Pimenta, 
Phys. Rev. B {\bf 76}, 201401(R) (2007).

\bibitem{HJTS} E. A. Henriksen, Z. Jiang, L.-C. Tung, M. E. Schwartz, 
M. Takita, Y.-J.Wang, P. Kim, and H. L. Stormer, 
Phys. Rev. Lett. {\bf 100}, 087403 (2008).

\bibitem{OFBB} M. Orlita, C. Faugeras, J. Borysiuk, J. M. Baranowski,
W. Strupi\'nski, M. Sprinkle, C. Berger, W. A. de Heer, 
D. M. Basko, G. Martinez, and M. Potemski, 
Phys. Rev. B {\bf 83}, 125302 (2011).

\bibitem{SMPB} M. L. Sadowski, G. Martinez, M. Potemski, C. Berger, 
and W. A. de Heer, 
Phys. Rev. Lett. {\bf 97}, 266405 (2006).

\bibitem{EGMM} 
D.~C. Elias, R.~V. Gorbachev, A.~S. Mayorov, S. V. Morozov, A.~A. Zhukov, P. Blake,
L.~A. Ponomarenko, I.~V. Grigorieva, K.~S.~Novoselov, F.~Guinea, and A.~K.~Geim, 
Nat. Phys. {\bf 7}, 701 (2011).

\bibitem{FBNH} C. Faugeras, S. Berciaud, P. Leszczynski, Y. Henni, K. Nogajewski, 
M. Orlita, T. Taniguchi, K. Watanabe, C. Forsythe, P. Kim, R. Jalil,  
A. K. Geim, D. M. Basko, and M. Potemski,
Phys. Rev. Lett. {\bf 114}, 126804 (2015).


\bibitem{HuntYY} B. Hunt, J. D. Sanchez-Yamagishi, A. F. Young, 
M. Yankowitz, B. J. Leroy, K. Watanabe, T. Taniguchi, P. Moon, 
M. Koshino, P. Jarillo-Herrero, and R. C. Ashoori, 
Science {\bf 340}, 1427 (2013).


\bibitem{CSYL} Z.-G. Chen, Z. Shi, W. Yang, X. Lu, Y. Lai, H. Yan, 
F. Wang, G. Zhang, and Z. Li, 
Nat. Commun. {\bf 5}, 4461 (2014).


\bibitem{WoodsBE} C.~R.~Woods, L.~Britnell, A.~Eckmann, R.~S.~Ma, J. C. Lu, 
H. M. Guo, X. Lin, G. L. Yu, Y. Cao, R. V. Gorbachev, A. V. Kretinin, J. Park, 
L. A. Ponomarenko, M. I. Katsnelson,
Yu. N. Gornostyrev, K. Watanabe, T. Taniguchi, C. Casiraghi, H-J. Gao, A. K. Geim,
and K. S. Novoselov, 
Nat. Phys. {\bf 10}, 451 (2014).


\bibitem{RZTW} B. J. Russell, B. Zhou, T. Taniguchi, K. Watanabe, and E. A. Henriksen,
Phys. Rev. Lett. {\bf 120}, 047401 (2018).


\bibitem{MOG} A. H. MacDonald, H.~C.~A. Oji, and S. M. Girvin, 
Phys. Rev. Lett. {\bf 55}, 2208 (1985).

\bibitem{GMP} S. M. Girvin, A. H. MacDonald, and P. M. Platzman,
Phys. Rev. B {\bf 33}, 2481 (1986).

\bibitem{MZ}  A. H. MacDonald and S.-C. Zhang,
Phys. Rev. B {\bf 49}, 17208 (1994).

\bibitem{KS_sma} K. Shizuya, Int.~J.~Mod.~Phys. B {\bf 31}, 1750176 (2017).

\bibitem{Semenoff}  G. W. Semenoff, Phys. Rev. Lett. {\bf 53}, 2449 (1984).

\bibitem{KS_screening} K. Shizuya, Phys. Rev. B {\bf 75}, 245417 (2007).

\bibitem{fn_comp} 
This sum rule in general holds for the eigenmodes of the one-body Hamiltonian;
for a proof see, e.g., K. Shizuya, Phys. Rev. B {\bf 87}, 085413 (2013).  


\bibitem{fn_ehconj} 
For a state consisting of  
four ($= 2_{\rm spin} \times 2_{\rm valley})$ Landau levels with $n_{\rm f}^{(i)}$,
the filling factor is written as $\nu = 2+ \sum_{i} n_{\rm f}^{(i)}$;
on setting $n_{\rm f}^{(i)} \rightarrow -n_{\rm f}^{(i)} -1$, $\nu$ turns into $-\nu$.


\bibitem{GKBK}  G. Giovannetti, P. A. Khomyakov, G. Brocks, P. J. Kelly, 
and J. van den Brink, 
Phys. Rev. B {\bf 76}, 073103 (2007).

\bibitem{SGS} P. San-Jose, A. Guti\'errez-Rubio, M. Sturla, and F. Guinea,
Phys. Rev. B {\bf 90}, 075428 (2014).

\bibitem{JDMA} J. Jung, A. M. DaSilva, A. H. MacDonald, and S. Adam,
Nat. Commun. {\bf 6}, 6308 (2015).

\bibitem{fn_suppl} The Supplemental Material of Ref.~\cite{RZTW} presents
some more $(v^{\rm eff}\ vs\ \nu)$ data at $B = (5,11)$T.
In the $B=11$T data $v^{\rm eff}$ shows a minimum at $\nu=4$ for $T_{2}$  
while no such minima are seen in the $B=5$T data.

\bibitem{SZL} A. A. Sokolik, A. D. Zabolotskiy, and Y. E. Lozovik, 
 Phys. Rev. B {\bf 95}, 125402 (2017).

\end{thebibliography}
\end{document}